\documentclass[conference]{IEEEtran}

\pdfoutput=1
\usepackage{url}
\usepackage{listings}
\usepackage{xcolor}
\definecolor{lightgray}{rgb}{.9,.9,.9}
\definecolor{purple}{rgb}{0.65, 0.12, 0.82}
\definecolor{darkgreen}{rgb}{0.00, 0.50, 0.00}
\usepackage{graphicx}
\usepackage{caption}
\usepackage{subcaption}
\graphicspath{ {figs/} }

\let\labelindent\relax 
\usepackage{enumitem} 
\usepackage{cleveref}
\usepackage{cite}

\usepackage{balance}

\usepackage[normalem]{ulem}

\usepackage{mdframed}
\usepackage{tabularx}
\usepackage{tablefootnote}

\newenvironment{nonbreakablebox}{\par\noindent\begin{minipage}{\linewidth}}{\end{minipage}\par}

\newcommand{\asm}[1]{{\tt\normalsize#1}}

\newcommand{\etal}[1]{#1 \emph{et al.}} 

\lstdefinestyle{customc}{
    belowcaptionskip=1\baselineskip,
    breaklines=true,
    frame=L,
    xleftmargin=\parindent,
    language=C,
    showstringspaces=false,
    basicstyle=\footnotesize\ttfamily,
    keywordstyle=\bfseries\color{green!40!black},
    commentstyle=\itshape\color{purple!40!black},
    identifierstyle=\color{blue},
    stringstyle=\color{orange},
}
\lstnewenvironment{ccode}[1][]{
    \lstset{
        style=customc,
        frame=single,
        #1
    }
}
{}

\lstdefinelanguage
[x64]{Assembler}     
[x86masm]{Assembler} 
{morekeywords={CDQE,CQO,CMPSQ,CMPXCHG16B,JRCXZ,LODSQ,MOVSXD, %
        POPFQ,PUSHFQ,SCASQ,STOSQ,IRETQ,RDTSCP,SWAPGS, %
        rax,rdx,rcx,rbx,rsi,rdi,rsp,rbp, %
        cpuid,rdtsc,%
        r8,r8d,r8w,r8b,r9,r9d,r9w,r9b, %
        r10,r10d,r10w,r10b,r11,r11d,r11w,r11b, %
        r12,r12d,r12w,r12b,r13,r13d,r13w,r13b, %
        r14,r14d,r14w,r14b,r15,r15d,r15w,r15b}} 

\lstdefinestyle{customasm}{
    keywords={cpuid, rdtsc, rdtscp},
    belowcaptionskip=1\baselineskip,
    frame=L,
    xleftmargin=\parindent,
    language=[x64]Assembler,
    basicstyle=\small\ttfamily,
    keepspaces=true,
    commentstyle=\itshape\color{purple!40!black},
    numbers=left,
    numberstyle=\footnotesize\color{gray},
    numbersep=5pt
}
\lstnewenvironment{asmcode}[1][]{
    \lstset{
        style=customasm,
        basicstyle=\ttfamily\small,
        frame=single,
        #1
    }
}
{}

\lstdefinelanguage{JavaScript}{
keywords={break, case, catch, continue, debugger, default, delete, do, else, false, finally, for, function, if, in, instanceof, new, null, return, switch, this, throw, true, try, typeof, var, void, while, with},
    morecomment=[l]{//},
    morecomment=[s]{/*}{*/},
    morestring=[b]',
    morestring=[b]",
    ndkeywords={class, export, boolean, throw, implements, import, this},
    keywordstyle=\color{blue}\bfseries,
    ndkeywordstyle=\color{darkgray}\bfseries,
    identifierstyle=\color{black},
    commentstyle=\color{purple}\ttfamily,
    stringstyle=\color{red}\ttfamily,
    sensitive=true
}

\begin{document}
\title{\textsc{Speculose}: Analyzing the Security Implications \\of Speculative Execution in CPUs}

\author{\IEEEauthorblockN{Giorgi Maisuradze}
    \IEEEauthorblockA{CISPA, Saarland University\\
        Saarland Informatics Campus\\
        giorgi.maisuradze@cispa.saarland}
    \and
    \IEEEauthorblockN{Christian Rossow}
    \IEEEauthorblockA{CISPA, Saarland University\\
        Saarland Informatics Campus\\
        rossow@cispa.saarland}}

\maketitle

\begin{abstract}
Whenever modern CPUs encounter a conditional branch for which the condition cannot be evaluated yet, they predict the likely branch target and \emph{speculatively} execute code.
Such pipelining is key to optimizing runtime performance and is incorporated in CPUs for more than 15 years.
In this paper, to the best of our knowledge, we are the first to study the inner workings and the security implications of such speculative execution.
We revisit the assumption that speculatively executed code leaves no traces in case it is not committed.
We reveal several measurable side effects that allow adversaries to enumerate mapped memory pages and to read arbitrary memory---all using only speculated code that was never fully executed.
To demonstrate the practicality of such attacks, we show how a user-space adversary can probe for kernel pages to reliably break kernel-level ASLR in Linux in under three seconds and reduce the Windows 10 KASLR entropy by 18~bits in less than a second.
\end{abstract}

\noindent
\textbf{Disclaimer:}
This work on speculative execution was conducted independently from other research groups and was submitted to IEEE S\&P '17 in October 2017.
Any techniques and experiments presented in this paper predate the public disclosure of attacks that became known as Meltdown~\cite{meltdown} and Spectre~\cite{spectre} and that were released begin-January 2018.

\section{Introduction}

Being at the core of any computer system, CPUs have always strived for maximum execution efficiency.
Several hardware-based efforts have been undertaken to increase CPU performance by higher clock frequencies, increasing the number of cores, or adding more cache levels.
Orthogonal to such developments, vendors have long invested in logical optimization techniques, such as complex cache eviction algorithms, branch predictors or instruction reordering.
These developments made clear that CPUs do not represent hardware-only components any longer.
Yet our level of understanding of the \emph{algorithmic}, i.e., software-side aspects of CPUs is in its infancy.
Given that many CPU design details remain corporate secrets, it requires tedious reverse engineering attempts to understand the inner workings of CPUs~\cite{opcode_reverse_x86}.

In this paper, we argue that this is a necessary direction of research and investigate the security implications of one of the core logical optimization techniques that is ubiquitous in modern CPUs: \emph{speculative execution}.
Whenever facing a conditional branch for which the outcome is not yet known, instead of waiting (stalling), CPUs usually speculate one of the branch target.
This way, CPUs can still fully leverage their instruction pipelines.
They predict the outcome of conditional branches and follow the more likely branch target to continue execution.
Upon visiting a particular branch for the first time, CPUs use \emph{static} prediction, which usually guesses that backward jumps (common in loops to repeat the loop body) are taken, whereas forward jumps fall through (common in loops so as not to abort the loop).
Over time, the CPU will learn the likely branch target and then uses \emph{dynamic} prediction to take the more likely target.
When a CPU discovers that it mispredicted a branch, it will roll back the speculated instructions and their results.
Despite this risk of mispredictions, speculative execution has significantly sped up CPU performance and is part of most modern CPUs of popular vendors Intel, AMD or ARM-licensed productions.

To the best of our knowledge, we are the first to analyze the security implications of speculative execution.
So far, the only known drawback of speculative execution is a slightly higher energy consumption due to non-committed instructions~\cite{speculation_powerreduction}.
As we show, its drawbacks go far beyond reduced energy efficiency.
Our analysis follows the observation that CPUs, when executing code speculatively, might leak data from the speculated execution branch, although this code would never have been executed in a non-speculative world.
Ideally, speculated code should not change the CPU state unless it is committed at a later stage (e.g., because the predicted branch target was confirmed).
We analyze how an adversary might undermine this assumption by causing measurable side effects during speculative execution.
We find that at least two feedback channels exist to leak data from speculative execution, even if the speculated code, and thus its results, are \emph{never} committed.
Whereas one side channel uses our observation that speculated code can change cache states, the other side channel observes differences in the time it takes to flush the instruction pipeline.

With these techniques at hand, we then analyze the security implications of the possibility to leak data from within speculative execution.
We first show how a user-space attacker can abuse speculative execution to reliably read arbitrary user-space memory.
While this sounds boring at first, we then also discuss how that might help an attacker to access memory regions that are guarded by conditionals, such as in sandboxes.
We then analyze if an unprivileged user can use speculation to read even kernel memory.
We show that this attack is fortunately not possible.
That is, speculative execution protects against invalid reads in that the results of access-violating reads (e.g., from user to kernel) are zeroed.

This observation, however, leads us to discover a severe side channel that allows one to distinguish between mapped and unmapped kernel pages.
In stark contrast to access-violating memory reads (which are zeroed), page faults (i.e., accesses to non-mapped kernel pages) stall the speculative execution.
We show that an attacker can use this distinction to reliably and efficiently determine whether a virtual memory page is mapped.
This effectively undermines a fundamental assumption of kernel-based Address Space Layout Randomization (KASLR) designs~\cite{linux_kaslr} present in modern OSes (Windows 8.x+, Linux kernel 4.4+, iOS 4.3+, Android 8.x):
KASLR's foremost goal is to hide the location of the kernel image, which is easily broken with speculative execution.
Access violations in speculative execution---in contrast to violations in non-speculated execution---do not cause program crashes due to segmentation faults, allowing one to easily repeat checks in multiple memory ranges.
In our experiments, we use commodity hardware to show that one can reliably break KASLR on Ubuntu 16.04 with kernel 4.13.0 in less than three seconds.

\vspace{0.5em}
\noindent
In this paper, we provide the following contributions:

\begin{itemize}[noitemsep,nolistsep]
    \item To the best of our knowledge, we are the first to explore the internals of speculative execution in modern CPUs.
    We reveal details of branch predictors and speculative execution in general.
    Furthermore, we propose two feedback channels that allow us to transfer data from speculation to normal execution.
	We evaluate these primitives on five Intel CPU architectures, ranging from models from 2004 to recent CPU architectures such as Intel Skylake.
    
    \item Based on these new primitives, we discuss potential security implications of speculative execution.
    We first present how to read arbitrary user-mode memory from inside speculative execution, which may be useful to read beyond software-enforced memory boundaries.
    We then extend this scheme to a (failed) attempt to read arbitrary kernel memory from user space.
    
    \item We discover a severe side channel in Intel's speculative execution engine that allows us to distinguish between a mapped and a non-mapped kernel page.
    We show how we can leverage this concept to break KASLR implementations of modern operating systems, prototyping it against the Linux kernel 4.13 and Windows 10.

	\item We discuss potential countermeasures against security degradation caused by speculative execution, ranging from hardware- and software- to compiler-assisted attempts to fix the discovered weaknesses.
\end{itemize}

\section{Analysis of Speculative Execution in x86}
\label{sec:background}

In this section, we first provide an overview of the basic x86 architecture in general.
We then introduce the concept of speculative execution and its implementation details in the recent x86 microarchitectures such as Haswell and Skylake.
To understand the inner workings of speculative execution, this section also describes the details of branch prediction techniques that are used by modern processors.

These inner details allow for an attack that abuses speculative execution to break kernel-level Address Space Layout Randomization (KASLR).
This section therefore also introduces KASLR, an in-kernel defense mechanism, deployed in most modern operating systems.
To add an emphasis the importance of KASLR, we will show a wide range of kernel attacks that are possible in the absence of KASLR.
Later, in \Cref{sec:attack}, we will combine branch prediction and speculative execution to anticipate the speculatively executed path after a conditional branch.
This will be a key to executing arbitrary code speculatively, which we will use in our attack to remove the randomness introduced by KASLR.

\subsection{Generic x86 architecture}

Despite being a CISC (Complex Instruction Set Computing) architecture, x86\footnote{Note that we will refer to both IA-32 (32-bit CPUs) and AMD64/Intel 64 (64-bit CPUs) as x86 for simplicity.} constitutes \emph{the} prevalent architecture used for desktop and server environments.
Extensive optimizations are among the reasons for x86's popularity.

Realizing the benefits of RISC (Reduced Instruction Set Computing) architectures, both Intel and AMD have switched to RISC-like architecture under the hood.
That is, although still providing a complex instruction set to programmers, internally they are translated into sequences of simpler RISC-like instructions.
These high-level and low-level instructions are usually called macro-OPs and micro-OPs, respectively.
This construction brings all the benefits of RISC architecture and allows for simpler constructions and better optimizations, while programmers can still use a rich CISC instruction set.
An example of translating macro-OP into micro-OP is shown in the following:
\begin{asmcode}
                    load  t1, [rax]
add [rax], 1   =>   add   t1, 1
                    store [rax], t1    
\end{asmcode}

Using micro-OPs requires way less circuitry in the CPU for their implementation.
For example, similar micro-OPs can be grouped together into execution units (also called ports).
Such ports allow micro-OPs from different groups to be executed in parallel, given that one does not depend on the results of another.
The following is an example that can trivially be run in parallel once converted into micro-OPs:
\begin{asmcode}
mov [rax], 1
add  rbx,  2
\end{asmcode}
Both instructions can be executed in parallel.
Whereas instruction \#1 is executed in the CPU's memory unit, instruction \#2 is computed in the CPU's Arithmetic Logic Unit (ALU)---effectively increasing the throughput of the CPU.
Based on the availability of their corresponding execution ports and the source data, each of these instructions could also be executed out of order.
In the previous example, the CPU could easily swap the order of addition and memory read, as there are no dependencies between them.
Such reordering frequently allows optimizing a sequence of instructions for speed and data locality.
Given the possibility of data dependencies in general, an execution cannot be done entirely out-of-order, though.
Otherwise, such reordering would create data hazards and possibly even result in entirely different executions that do not reflect the macro-OP specification.
To capture the complexity of such reordering, CPUs include another unit that maintains the consistency of the executed instructions.

\begin{figure}[t]
    \includegraphics[trim=100 60 0 0,width=1.13\columnwidth]{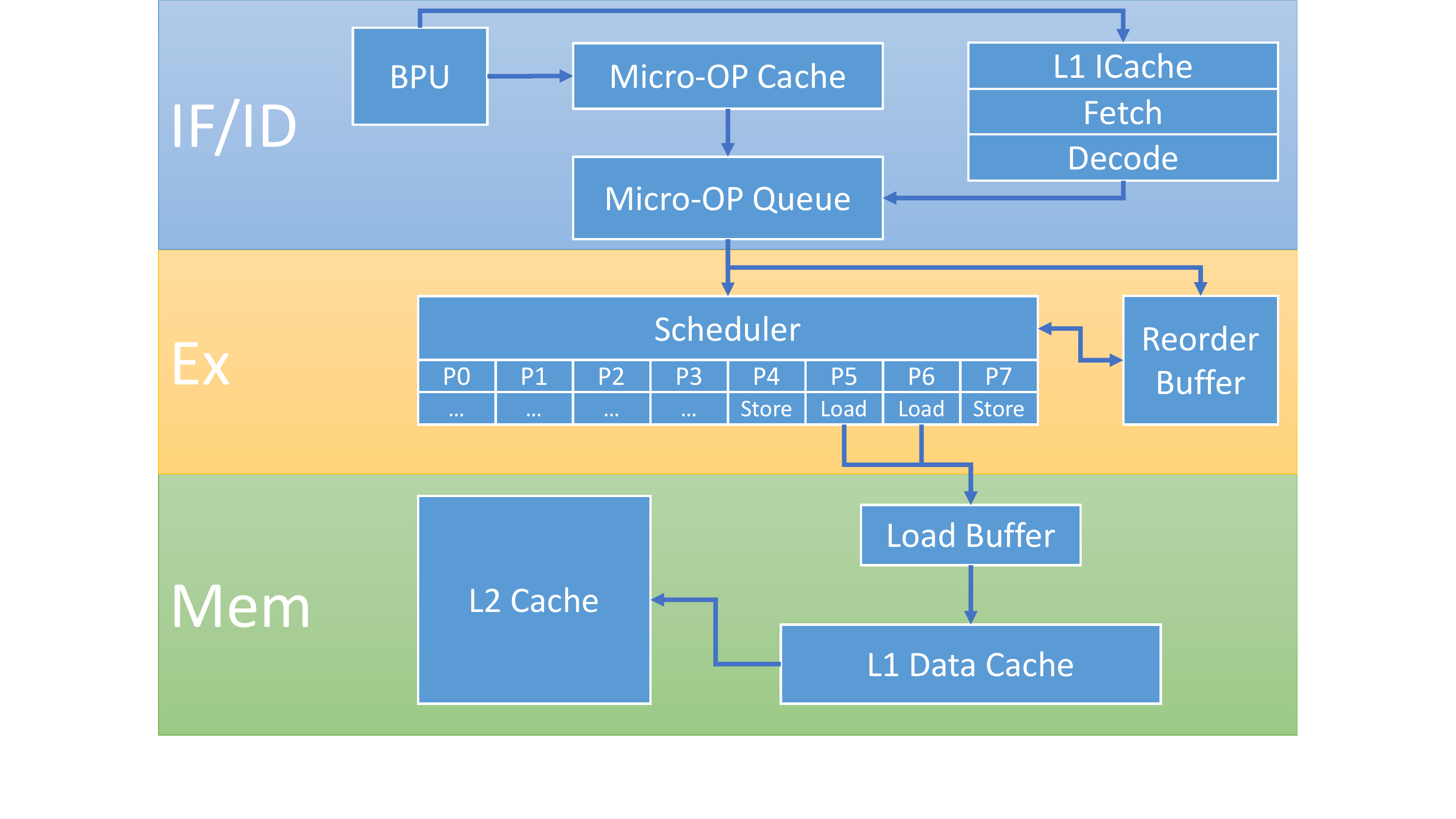}
    \captionsetup{justification=centering}
    \caption{CPU Core Pipeline Functionality of the Haswell Microarchitecture~\cite{intel_optimization}.}\label{fig:haswell_simple}
\end{figure}

\subsubsection{Execution Units in Modern x86 CPUs}
As software changes over time, the underlying logic and hardware setup of CPUs is also subject to change.
CPU vendors usually maintain microarchitectures to summarize various CPUs of the same generation, possibly varying in terms of clock speeds or number of cores.
We will now describe the basic x86 CPU architecture based on the example of the Haswell microarchitecture, which is a recent Intel x86 microarchitecture and one of the microarchitectures that we will use in our experiments.
Although the inner details differ from one microarchitecture to another, the basic building blocks (especially the ones that we use) remain mostly the same.

In \Cref{fig:haswell_simple}, we outline the most important parts of CPU's pipeline, based on the Haswell microarchitecture~\cite{intel_optimization}.
We split the whole pipeline into three parts: (i) Instruction fetch/decode, (ii) Execution, and (iii) Memory.

\textbf{Fetch/Decode Unit:}
The fetch/decode unit is responsible for fetching the forthcoming (i.e., to-be-executed) instructions and preparing them for execution.
There are different places where these instructions might be coming from.
The fastest of these places is a micro-OP cache, which is a storage of instructions that are already decoded into their corresponding micro-OPs.
If there is a miss in this cache, then the instruction has to be retrieved from one of the caches in the hierarchy (L1-I, L2, L3, L4), or ultimately, if none of the caches contain the data, from main memory.
The fetched instructions are then decoded and passed to the execution unit as well as to the micro-OP cache.
The details of how this is handled are not important for the remainder of our paper and we refer the interested reader to the respective CPU vendor documentation~\cite{intel_optimization}.
One crucial part to note, though, is that the address of the next instruction to fetch is controlled by the Branch Prediction Unit (BPU).

\textbf{Execution:}
After decoding the instruction into the corresponding micro-OPs, they are put on the instruction dispatch queue to prepare for execution.
Micro-OPs on the queue are checked for dependencies, and the ones with all source operands resolved are scheduled to one of 8 (in Haswell; varies in other architectures) execution ports.
Further, the scheduled micro-OPs are added to the reorder buffer (ROB), which contains the sequentially consistent view of all executed instructions.
Abstracting away from its details, it is sufficient for us to imagine the ROB as a unit that keeps micro-OPs in order while executing them out of order.
Usually, the ROB is implemented as a ring buffer, with its head pointing to the newest micro-OP and its tail to the oldest not-yet-committed operation.
\emph{Committing} a micro-OP means to reflect the changes that it made to the machine state.
Committing micro-OPs is done in order, i.e., a micro-OP can be committed only after it is done executing and all instructions before it are already committed.
In contrast, however, finishing the computation does not directly mean that the micro-OP can be committed.
A relevant example in our case is speculative execution, in which case a micro-OP can only be committed after the outcome of the speculation is known (i.e., after knowing that the speculated path was correctly predicted).
This implies that CPUs have to maintain micro-OPs uncommitted until they can be committed.
CPUs typically employ a maximum number of uncommitted/speculated instructions which aligns to the number of ROB entries (192 in Haswell).
This limit might be further reduced by the availability of other resources, such as physical registers or available execution ports.

\textbf{Memory Unit:}
The third part of the execution pipeline, the memory unit, is used for micro-OPs that access memory.
Being the bottleneck in most executions, CPU designers pay special attention to this unit to improve its efficiency.
That is the reason why in modern architectures we have a hierarchy of memory caches, each containing different sizes of buffers with varying access speeds.
Haswell's memory hierarchy, for example, comprises Load and Store Buffers (72 and 42 entries in Haswell, respectively), followed by various levels of caches (L1-I/L1-D, L2, L3, L4), and finally the main memory.
Each load operation is done in the following way:
\begin{enumerate}
	\item When a \texttt{load} micro-OP is scheduled, a load buffer entry is allocated that will hold the loaded data until it is committed.
	
	\item To retrieve the data, the L1 data cache is queried. In case of a cache hit, the data will be loaded into the load buffer entry of the corresponding micro-OP.
	
	\item In the case of an L1 cache miss, the data will be queried from the higher cache hierarchies or, ultimately, from the main memory.
\end{enumerate}
Executing a memory load involves two buffers.
First, a load buffer entry is allocated which will be tied to the corresponding \texttt{load} micro-OP, and will hold the loaded data after it is returned from the L1 data cache.
Second, in the case of a cache miss, the L1 data cache has to load the missing data from the higher memory hierarchy.
To service multiple cache misses in parallel, the L1 data cache has its own buffer consisting of memory accesses that have to be resolved.
Haswell, e.g., has a 32-entry buffer and can handle 32 cache misses\footnote{Note that in reality even cache hits will allocate entries in the buffer; however, as cache hits are serviced instantaneously, they can be ignored.} in parallel.

Executing \texttt{load} micro-OPs speculatively means that new load buffer entries will be allocated for each of them; however, they will not be freed until the CPU is done speculating. 
As commits are not carried out until the result of speculation are known, the loaded data cannot reach the architectural register and cannot be discarded yet.
This means that while speculating, the maximum number of load instructions that can be executed is limited by the size of the load buffer.
After reaching the limit, the execution of subsequent load micro-OPs will be stalled.
Given the generous sizes of load buffers in recent microarchitectures (72 entries in Haswell), for most programs this will not be an issue.
However, we will use this limitation as an advantage when measuring the access times of mapped/unmapped memory accesses in \Cref{sec:buffer-exhaustion}.
Also worth noting here is the case of cache misses in the L1 data cache.
As we previously mentioned, there is a limited number of ongoing cache misses that can be handled by the L1 data cache (32 in Haswell).
This, combined with the fact that faults cannot be handled speculatively, means that after having 32 faulty cache misses (e.g., cache miss resulting in page fault), all the subsequent memory operations will be stalled.

\subsubsection{Speculative Execution}
Speculative execution is an important optimization carried out by most CPU architectures.
Programs, generally, have many conditional branches.
Evaluating the condition in a conditional branch might take some time (e.g., reading data from memory, which can take hundreds of cycles if it comes from main memory.
Instead of waiting for a branch condition to be evaluated, the CPU speculates on the outcome and starts speculatively executing the more likely execution path.
Once the condition for the branch can be evaluated, this results in two possible scenarios:
    \begin{enumerate}
        \item The prediction was correct $\rightarrow$ The speculatively executed instructions were correct and thus will now be committed.
        \item The prediction was wrong $\rightarrow$ The speculatively executed instructions are flushed and the correct path is executed from the beginning (non-speculatively).
    \end{enumerate}

Optimizing the correctness of branch prediction is thus of utmost importance for performance.
Perfect prediction means that the CPU never has to wait for slow branches and will always execute the correct path.
In contrast, mispredictions penalize the execution by forcing the CPU to flush all the progress it has made since the speculation and start over from scratch.

\subsection{Branch Prediction}
\label{sec:branch-brediction}

As the efficacy of branch predictors directly influences the performance of CPUs, hardware vendors have improved predictors over time.
Therefore, branch predictors not only differ from vendor to vendor, but also between different generations of microarchitectures (e.g., Haswell vs, Skylake).
The basic principle is the same: branch predictors try to foresee the outcome of conditional branches.
There are generally two types of predictors, static and dynamic.
When a CPU sees a conditional branch for the first time, it has to resort to a static predictor to guess the outcome based on simple expert-knowledge heuristics.
In contrast, if the branch has been executed previously, a dynamic predictor can make more informed decisions by looking at previous branch outcomes.

The simplest example of a static predictor can be predicting any branch to be either always taken or never taken.
However, intuitively, these predictors are not highly precise.
The most widely used static predictor, that also exists in most modern CPUs, is called BTFNT (backwards taken forwards not taken).
As its name suggests, this static predictor predicts forward branches not to be taken and backwards to be taken.
The reasoning is that loops (i.e., backward jumps) are usually taken more than once, and the first case of conditionals (i.e., fall-through cases of forward jumps) are more likely to be taken.
The success rate of the predictor can be further improved by the compiler, which aligns the cases of conditionals according to the static predictions of the underlying hardware.

The simplest example of a dynamic branch predictor is a one-bit history predictor, which stores a single bit for each branch instruction.
This bit denotes whether the branch was taken or not in the previous execution.
Most CPUs are believed to\footnote{Branch predictor details are not part of official vendor documentation.} have 2-bit predictors (also called 2-bit saturating counters)~\cite{branch_prediction_stragegies}.
These two bits represent a counter that is incremented every time the branch is taken and decremented otherwise.
To decide the outcome of the predictor, the values of the counter \{0, 1, 2, 3\} are mapped to decisions \{strongly not-taken, weakly not-taken, weakly taken, strongly taken\}.
Apart from simple history matching, dynamic predictors can also detect cycles or nested branches, e.g., by using local or global history bits or even combining them together.
However, dynamic branch predictors are out of scope of this paper and therefore, we will not go into more detail here.

In the following, we will use static predictors, as they show a relatively coherent behavior across different CPUs, making it possible to reliably anticipate their outcome.
For example, if a CPU sees a branch at address \asm{A} with a forward conditional branch for the first time, the static branch predictor will predict the branch to be not taken (according to BTFNT).
In principle, the same thing can also be done on dynamic predictors by training them at the beginning and then taking the opposite branch of what the \textit{trained} predictor expects.

\subsection{Kernel ASLR}
\label{sec:background-KASLR}
With this basic CPU knowledge, we now turn to ASLR, a defense mechanism whose underlying assumptions an adversary can undermine with speculative execution.
Address Space Layout Randomization (ASLR)~\cite{pax:aslr} is a widely deployed defense technique against code-reuse attacks.
In user space, ASLR randomizes the base addresses of the program's memory segments, thus preventing the attacker from knowing or predicting the addresses of gadgets that she needs for code-reuse attacks.
ASLR is applied to a program at load-time and whenever a program requests a new memory allocation.
ASLR raises the bar for adversaries in doing successful exploitation, and thus is supported by all modern operating systems.

The recent increase of attack targets towards the kernel side~\cite{return-oriented-rootkits,pikit,ret2dir_rethink_kernel_isolation, rowhammer, dirty-cow} motivated the operating system developers to create similar countermeasures against code-reuse attacks for kernel code.
This resulted in creating KASLR (Kernel ASLR), which is an ASLR implementation for the kernel, i.e., randomizing the kernel's image base address as well as its loaded modules/drivers and allocated data.
While historically the kernel image was loaded at a fixed address (e.g., \texttt{0xffffffff80000000} in Linux), with KASLR, the kernel is placed at a different address at boot time.
KASLR is supported by all major operating systems, Windows (from Vista), OSX (from Mountain Lion 10.8) and Linux (from 3.14, and enabled by default from 4.12).
KASLR is meant to (1) protect against remote attackers, and (2) as a defense against privilege escalation attacks from local adversaries.
In our KASLR derandomization attack, we assume the latter case, i.e., when the attacker tries to reveal the location of critical kernel functions in order to use this in a subsequent exploitation.

\textbf{Windows KASLR:}
KASLR implementations differ between OSes, or even among different versions of the same OS.
In general, at every boot, the kernel image and the loaded modules/drivers are allocated at random addresses.
In Windows~10 (version 1709), there is a dedicated kernel memory region in the range from \texttt{0xfffff8\textbf{00000}00000} to \texttt{0xfffff8\textbf{80000}00000}, containing allocation slots of 2MiB each (i.e., large pages).
KASLR will then randomly assign these slots to the kernel image and to the loaded modules, resulting in ‭262144‬ possible slots where the kernel image can reside, i.e., 18 bits of randomness.

\textbf{Linux KASLR:} 
In Linux, KASLR randomly chooses any page from which to start loading the image.
By default, the size of this page is 2MiB, and the range in which the randomization can happen is  from \texttt{0xffffffff\textbf{800}00000} to \texttt{0xffffffff\textbf{c00}00000}.
This gives at most 512 possible kernel base image start pages (minus the image size itself), i.e., 9 bits of randomness.
The dedicated memory range of the kernel image is followed by a memory region for kernel modules.
It starts at \texttt{0xffffffff\textbf{c0000}000} and can go as far as \texttt{0xffffffff\textbf{fd200}000}.
However, only the load offset is randomized, i.e., the first loaded module will be loaded at a randomly chosen 4KiB page from the range [1,1024].
All consecutively loaded modules will then follow the first one, giving in total 10 bits of randomness.
\section{Threat Model}
This section details our threat model and assumptions we make about the attack environment.
Our threat model assumes a local attacker and is in accordance with the threat model of other attacks against KASLR~\cite{kaslr_break_tsx,kaslr_break_prefetch,kaslr_break_branchpredictor,kaslr_break_practicaltiming}.
More specifically, we envision the following scenario:

\begin{enumerate}
    \item An attacker can already execute arbitrary code on the victim's system in \emph{user mode} without elevated privileges.
    This means that either the attacker is running her own executable as a local user on the machine, or can inject arbitrary code in an already running user program.
    
    \item The attacker knows an exploitable kernel vulnerability and aims to abuse this vulnerability to elevate her privileges.
    For example, a vulnerability could allow her to hijack the kernel process's control flow (if the attacker knows the precise memory location of the kernel).
    
    \item We assume the OS deploys KASLR that mitigates such exploitation attempts by randomizing the kernel's image base address.
    This implies that there is no other way to leak address space information from the kernel that would otherwise undermine KASLR.

    \item Given that a na\"ive trial-and-error search would crash the kernel, we assume that the attacker cannot just brute-force kernel addresses.
\end{enumerate}
\section{Potential Abuses of Speculative Execution}
\label{sec:attack}
A key observation of our analysis in \Cref{sec:background} is that in case of a branch misprediction, the CPU has been executing the set of instructions that it was actually not supposed to execute.
We will now outline techniques where an attacker can abuse speculative execution to leak certain information (e.g., memory content, mapped pages), ultimately compromising fundamental assumptions of KASLR in Linux and Windows.
First, we will provide an overview of the basic approach and show several code patterns that allow abuse of speculative execution as a potential side channel.
We will then discuss how an attacker could use these techniques to identify the memory region of a KASLR-protected kernel.

\subsection{Enforced Speculative Execution}
If we trick the CPU into a branch misprediction, it will speculatively execute code that was not actually meant to be executed.
But: \emph{How can we force the CPU to enter code parts that we want to execute speculatively?}
To execute a piece of code speculatively, we first have to ensure that the branch condition takes ``long'' to evaluate, and in the meanwhile code after the conditional jump is executed in parallel.
To slow down branch condition computation, we propose to fill up a port of the CPU with many computations that the condition has to wait for before it can evaluate.
For the remainder of this paper, we will use integer multiplications (\asm{imul reg32,imm32}) to compute an input for a conditional branch.
The \asm{imul} instruction multiplies register \asm{reg} by immediate \asm{imm} and stores the result in \asm{reg}.
In our example, several (we use 2048 in the remainder of this paper) subsequent \asm{imul} will fill the multiplication port.

In addition, we have to make sure that the condition and the speculated code are not scheduled on the same execution ports.
Otherwise, there will be no free execution units to run the speculated code.
For example, if we want to use memory reads (\asm{mov reg32,mem32}) within speculative execution, these instructions will be scheduled on memory load ports (5 \& 6 in Haswell).
While ports may slightly differ between architectures, they (and thus the described procedure) are relatively stable and well-documented in vendor manuals~\cite{intel_optimization}.

\begin{asmcode}[label=lst:simple-speculation,caption=Code pattern to trigger speculative execution via static prediction (BTFNT) of the conditional jump in line~5]
  imul    r9,  3    ; Repeat imul to fill
  ...               ;  the ALU's queue
  imul    r9,  3    ;  
  cmp     r9b, 3    ; Requires imul result  
  je      C1True    ;  and has to speculate 
  << ... >>         ; Code that is executed
  << ... >>         ;  speculatively
  jmp     Exit
C1True:
  << ... >>
Exit:
\end{asmcode}

\Cref{lst:simple-speculation} outlines the basic idea.
We first fill a port with many \asm{imul} instructions that depend upon each other (lines~\#1--\#3).
The compare instruction (line~\#4) has to wait for the result of this multiplication chain, while the CPU speculates that the conditional forward jump (line~\#5) is not taken.
This code pattern thus allows us to (speculatively) execute arbitrary code (placeholders in lines \#6--\#7) that would have never been executed without speculation.
In all our experiments, the initial value of \asm{r9} is 3.
Therefore, after multiplying it 2048 times by 3, the least significant byte of \asm{r9} will be $3^{2048+1} $ mod $256 = 3$, i.e., the outcome of the conditional jump (line~\#4) was mispredicted.

When executing this code several times, the dynamic branch predictor will kick in and correctly predict the outcome.
To avoid this we use ASLR, i.e., we re-run the program after each execution, which will re-randomize the base address of the executable and thus the address for which the prediction is stored.
This forces the static predictor to make decisions again.

If an OS does not randomize the base address (either lack of ASLR, or ASLR randomization as boot time, such as in Windows), we have to search for alternative solutions.
One (that we will use later) is writing a larger probe program that runs several speculative executions in batch and thus just needs to be called once.
This leaves dynamic prediction no chance, but allows executing long code speculatively.
Another solution would be to manually unlearn the prediction result.
This could, e.g., be done by running the program with inverted conditionals for several times before calling it with the actual conditionals---thereby enforcing the dynamic predictor to mispredict.

\subsection{Feedback Channel from Speculation}
\label{sec:backreport}
While we explained speculative execution can now be enforced, we will now discuss how any computations done in speculation can feed back data to the outside (non-speculated) execution.
We face a natural challenge.
As soon as the CPU would identify a misprediction, it would flush (revert) the mispredicted branch and would thus not commit its result.
That is, without special care, the outside world could not see any effects from speculative execution.
That means that we have to find a feedback channel for the data we want to leak during speculative execution.
That is, to reason about the outcome of the speculated code, we need to communicate back from the speculative execution to the non-speculated execution flow of the program.

Although ideally it is not supposed to, we find that speculative execution \emph{modifies} the CPU state in such a way that it becomes measurable in non-speculated execution.
This allows us to leak data from within speculation, although the explicit results of the speculated code have been flushed and are not observable.
As direct communication from speculative to regular execution is unfeasible, we have to think of potential side channels.
We describe two exemplary feedback channels that we can use for these goals in the following.

\subsubsection{Caching Side Channel Feedback}
\label{sec:caching_feedback}
Our first feedback channel reads memory inside speculative execution.
Such reads store the memory content in the L1 data cache, which can be detected by measuring its subsequent access time from normal execution.
This technique can be used to leak a single bit of information per cached line. 
In the remainder of this paper, we mainly use this approach, as shown in \Cref{lst:caching-feedback}.

\begin{asmcode}[label=lst:caching-feedback,caption=Code fragment for caching feedback channel]
  << ... >>       ; slow condition -> ZF=1
  jz  C1True
  mov rbx, [rax]  ; cache memory at rax
  jmp Exit
C1True:
  << ... >>       ; real execution
Exit:
  << measure [rax] access time >>
  << fast when cached, slow otherwise >>
\end{asmcode}

\subsubsection{Conditional Flushing Slowdown Feedback}
Our second feedback channel measures the time it takes to recover from mispredicted branches.
The core observation here is that the recovery time will depend on the state of the pipeline when flushing occurs.
For example, if we consider the code in \Cref{lst:flushing-feedback}, the time it takes to flush the pipeline will be less if we stop the execution (\asm{hlt}) from inside the speculative execution.
This is true even if the executed instructions are simple \asm{nop}s.
Note that, as we use caching as our main source of feedback channel, we have not investigated all possible types of such side-effect-incurring instructions.

\begin{asmcode}[label=lst:flushing-feedback,caption=Code for conditional flushing slowdown channel]
  << ... >>       ; slow condition -> ZF=1
  << Start Measurement >>
  jz  C1True
  nop             ; repeat
  ...             ; nops 
  nop             ; 128 times
  [hlt]           ; halt the execution
  jmp Exit
  C1True:
  << ... >>       ; real execution
  Exit:
  << End Measurement >>
\end{asmcode}

\subsection{Detecting Static Predictor Behavior}
\label{sec:detect_predictor}

We have described in \Cref{sec:background} that there are multiple strategies for static predictors.
The code mentioned in \Cref{lst:simple-speculation}, for example, assumes that the CPU statically predicts that forward jumps are not taken.
In the following, we outline two tests to reveal the forward jump prediction strategy for a specific CPU.
The first test aims to find out if the forward jumps are predicted to be taken or not.
The code in \Cref{lst:detect-predictor-FWD-NT} checks if forward jumps are predicted to be \emph{not} taken:
\begin{asmcode}[label=lst:detect-predictor-FWD-NT,caption=Detect forward-not-taken predictor]
  << ... >>          ; slow condition->ZF=1
  je      C1True     ; Condition is true
C1False:
  mov     rsi, [r11] ; r11 user address
  jmp     Exit
C1True:
  jmp     Exit
Exit:
  << measure [r11] access time >>
\end{asmcode}

In the above listed code, the conditional branch (line \#2) will eventually evaluate to true, i.e., it will jump to \texttt{C1True}.
That is, the repetition of \texttt{imul} before (line \#1) will fill the port such that the condition is not known and speculative execution will be started.
If the branch was predicted not to take the forward jump, then the memory read (line \#4) will be speculated, caching the value at \asm{r11}.
Checking the access time of memory at \asm{r11} (line \#9) reveals the decision, i.e., if access to \asm{r11} is faster than non-cached access, the predictor uses the fall-through case for forward jumps.

On the other hand, to check if forward jumps are predicted as taken, we can invert the above listed code by inverting the condition of the conditional jump:
\begin{asmcode}
  << ... >>          ; slow condition->ZF=1
  jne     C1True     ; Condition false
C1False:
  jmp     Exit
C1True:
  mov     rsi, [r11] ; r11 user address
  jmp     Exit
Exit:
  << measure [r11] access time >>
\end{asmcode}

Out of the CPUs that we tested, only the Sandy Bridge microarchitecture seems to be predicting forward jumps as taken.
Others speculate the fall-through case, which is the expected outcome.
Note that the prediction behavior is independent from the condition, e.g., both \asm{je} and \asm{jne} will be predicted the same (either both jump, or both fall through).
For simplicity, all following examples will assume that fall-through cases are speculated.
In our experiments (e.g., on Sandy Bridge), we will adjust the branch behavior accordingly.

\subsection{Arbitrary Memory Read}
\label{sec:arbitrary-memory-read}

We now try to abuse speculative execution to read arbitrary memory that should not be accessible during normal (committed) execution.
The basic idea is as follows: in speculative execution we check the memory content at a specified address, and report the result back using one of the feedback channels described in \Cref{sec:backreport}.
To leak an arbitrary byte in memory, we use several \asm{test} instructions to leak a byte bit-by-bit, i.e., it takes 8 checks to read a byte from memory.

\begin{asmcode}[label=lst:double-spec,caption=Arbitrary memory read via nested speculation]
  << ... >>           ; slow condition->ZF=1
  je      C1True      ; Taken
C1False: 
  ; --- START of SPECULATION ---
  mov     r10b, BYTE PTR[r10]
  test    r10b, 1     ; test 1st bit
  jz      C2True
C2False:                ; 1st bit is 1
  jmp     Exit        ; Lvl2 Speculation
C2True:                 ; 1st bit is 0
  mov     rsi, [r11]  ; cache mem @ r11
  jmp     Exit
  ; --- END of SPECULATION ---
C1True:
  << ... >>
Exit:
  ; if [r11] is cached
  ;   1st bit of [r10] = 1
  ; else
  ;   1st bit of [r10] = 0
\end{asmcode}

\Cref{lst:double-spec} shows the code listing that we can use to read one bit of arbitrary memory.
In this example, we want to read a value that is located at address \asm{r10}:
While the multiplication results are being evaluated (line \#1), \texttt{C1False} is executed speculatively.
Inside, we read a byte stored at the target address (\asm{r10}) into register \asm{r10b} (line~\#5).
Based on the least significant bit of the read value, we make the decision to either jump to \texttt{C2False} or \texttt{C2True}.
Until the memory at \asm{r10} (line \#5) is read, \texttt{C2False} (line \#8) will be run in \emph{nested} speculation---regardless of the pending test result (line \#6).
Assuming that the memory read (condition for second-level speculation) is faster than the sequence of multiplications (condition for first-level speculation)\footnote{We verify this assumption as part of our experiments.}, the outcome of the second branch condition will be known sooner than the first one.
Once the test result and thus the condition of the second conditional jump (line \#7) is known, the second-level speculation will be flushed.
The CPU would then start to execute the \emph{correct} jump target (line \#10).
Note that this does not flush the first-level speculation (that started in line \#3).
Therefore, \texttt{C2True}, which caches the feedback address \asm{r11} (line \#11), is executed only in the case that the read memory has 0 for its least significant bit.
The side effects of caching \asm{r11} can be easily detected from outside the speculative execution (line~\#16), when the multiplication result is finally known and the speculated paths are flushed.

By repeating this code snippet with different bit offsets to test, we can leak entire bytes or memory ranges.
Technically, we need to flush the feedback address \asm{r11} from the cache so that we can start fresh measurements, e.g., using \asm{clflush}.

\subsection{Arbitrary Kernel-Space Memory Read (But...)}
\label{sec:double-speculation-kernel}
Being able to read arbitrary user memory might have some interesting use cases, even for local attackers which seemingly already can read arbitrary code (for discussion see \Cref{sec:discussion}).
Following our threat model of a user-space attacker, the immediate follow-up question is now:
\emph{Can we even read memory in regions for which we do not have privileges?}

In our second experiment we thus aim to read privileged memory in kernel-land as an unprivileged user.
In other words, we want to check if privilege separation between user and kernel was taken into account also during speculative execution.
To this end, we ran the code from \Cref{lst:double-spec} with probe address pointing to a kernel address.
After running the code against various microarchitectures, much to our surprise, we saw that the code was being executed and the kernel memory was read.
However, the values that we read were \emph{not} the true values we expected.
More specifically, we observed that instead of getting the actual content of the memory, we would read 0s for all read kernel addresses.
This is good news in that an arbitrary kernel memory read is not possible from user space, which would be a complete disaster for security.

\begin{asmcode}[label=lst:double-speculation-kernel,caption=Checking if address \texttt{K} is mapped]
	<< ... >>           ; slow condition->ZF=1
	je      C1True     
	C1False:
	; --- START of SPECULATION ---
	cmp     [r10], 0    ; r10 = K
	jz      C2True
	C2False:              ; page is not mapped
	jmp     Exit
	C2True:               ; page is mapped
	mov     rsi, [r11]  ; r11 user address
	jmp     Exit
	; --- END of SPECULATION ---
	C1True:
	jmp     Exit
	Exit:
	<< measure [r11] access time >>
\end{asmcode}

Interestingly, however, our measurement results differed completely when we tried to read the values of kernel-land virtual addresses that did not map to any physical page.
In such a case, the execution of the corresponding micro-OP would stop, together with the micro-OPs that use the value read from it.
For example, if we read a non-mapped value into a register \asm{rax}, using \asm{rax} anywhere in subsequent instructions as a source operand will stall its execution.
This leads to a clear distinction between an access violation that results in a 0 being read, while in the case of a page fault (e.g., when the memory page is not mapped) the entire execution stalls.
An attacker can use this side channel to distinguish between mapped and unmapped kernel pages.
If a byte read from kernel is 0, the page was mapped; otherwise the page was not mapped.
To check if a kernel address \texttt{K} is mapped or not, we use the code from \Cref{lst:double-spec} with the exception that now we check for the memory content to be 0.
The assembly code for it is displayed in \Cref{lst:double-speculation-kernel} (\asm{r10} is \texttt{K}).

\noindent
We can further simplify this code by removing the second level conditional as shown in \Cref{lst:dependency-load-kernel}.
The observation here is that whenever we read something from a mapped kernel page we get a 0 back, and whenever the page is not mapped execution stalls.
Therefore, we can use the read value as an offset in a memory access that is caching the user-space address (line~\#6).
The offset in \asm{rax} is 0 in the case of a GPF and thus \asm{r11} will be cached.
In the case of a non-mapped page, the execution stalls after line~\#5.
Note that removing the register offset would remove the dependency between lines \#5 and \#6 and thus allow for instruction reordering, destroying the side channel.

\begin{asmcode}[label=lst:dependency-load-kernel,caption=Checking mapped pages via load dependency]
  << ... >>          ; slow condition->ZF=1
  je      C1True     
C1False:
  ; --- START of SPECULATION ---
  mov     rax, [r10]     ; r10 = K
  mov     rsi, [r11+rax] ; r11 user address
  jmp     Exit
  ; --- END of SPECULATION ---
C1True:
  jmp     Exit
Exit:
  << measure [r11] access time >>
\end{asmcode}

\subsection{Reasoning About the Observed Side Channel}
\label{sec:buffer-exhaustion}
Having this side channel at hand, it is not hard to see that schemes like KASLR that try to hide the presence of kernel pages are severely threatened.
One immediate question that came to our mind was:
\emph{Why does the CPU proceed with speculative execution in the case of access violations (thereby creating this side channel)?}
We expected the CPU to stall the execution whenever any fault occurs, be it either insufficient privileges or a page fault.
In practice, however, we see a clear difference.

In the following, we try to explain the underlying process in these two scenarios.
Due to the lack of documented details on CPU internals, this explanation has to be taken with a grain of salt.
The core problem is caused by the fact that faults cannot be handled by the memory unit during speculative execution, i.e., they are only dealt with at the commit stage.
In the case of kernel memory accesses, we have two types of faults.
A general protection fault (GPF) occurs if memory is mapped, but we do not have sufficient access rights to read it.
Second, a page fault (PF) occurs when memory is not mapped at all.
Given that memory units cannot handle faults speculatively (until later at commit stage), there are two different scenarios how these two faulty memory accesses are serviced.
In the case of GPF, the memory access is actually still carried out by the L1 data cache.
However, because of a privilege fault, the actual data is not put into a load buffer entry; instead the entry is marked as faulty to be handled later at commit stage.
Remember that each memory access allocates a new load buffer entry, which remains allocated until the micro-OP is not committed.
The observed value 0 in such cases can either be a default load buffer value when it is reserved for the micro-OP, or a value used by the L1 cache for faulty accesses.

In contrast, for PF, we have a cache miss that will never be resolved as the page is not mapped.
Therefore, the L1 data cache cannot return any value to the load buffer (because it does not have any), and it cannot finish the request (because it cannot handle faults while speculating).
Note that the L1 data cache has a limited number of entries for handling ongoing cache misses.
This means that the entry for handling the memory read will remain allocated for the entire speculation.
This will additionally stall the load buffer entry that waits for the result and any instructions that depend on the read value.

\subsection{Kernel-Probing Specific Feedback Channel}
\label{sec:probing-feedback}
We can leverage the fact that page faults stall the execution engine to design an even easier and faster way to determine whether a page is mapped.
For this we use the following observation: for non-mapped pages the bottleneck of the number of memory accesses is the size of ongoing loads in the L1 data cache (e.g., 32 in Haswell).
For mapped addresses, the bottleneck is the number of load buffer entries (e.g., 72 in Haswell).
This means that, if we have a non-mapped address \texttt{K}, doing \asm{mov rax,[K]} 32 times will stall the execution of following load instructions.
However, if address \texttt{K} is mapped, then the execution of loads will only stall after the CPU runs out of load buffer entries, i.e., after executing \asm{mov rax,[K]} 72 times in Haswell.
Therefore, any number of loads (\asm{mov rax,[K]}) between 32 and 71, followed by reading a user-space feedback address \texttt{U} (\asm{mov rsi,[U]}), will result in caching \texttt{U} if the page at address \texttt{K} is mapped.
\section{Evaluation}
\label{sec:evaluation}

We will now perform the described analyses on five x86 microarchitectures.
Each microarchitecture is a different iteration of implementation of CPU circuitry, thus affecting the execution process while still keeping the high-level semantics mostly the same.
To see how inner workings of CPUs may have changed, we gathered a wide selection of Intel CPUs with architectures ranging from the year 2004 to a recent one from 2015.
The complete list of the microarchitectures used in our experiments is shown in \Cref{tbl:uarchs-tested}.
Throughout the remaining paper, we will refer to each of the CPUs by its microarchitecture codename.

\subsection{Revealing Microarchitecture Details}

As a first experiment, we applied several tests described in \Cref{sec:attack} to reveal the inner workings of the different architectures.
That is, we experimentally reveal the static prediction strategy, the number of load buffer entries, and the maximum number of parallel cache misses.
\Cref{tbl:uarchs-properties} summarizes the outcomes of those measurements, showing that the various Intel architectures differ in their characteristics.
The general trend is that both the number of load buffer entries and parallel loads increased over the years.
For the static prediction pattern, there is a more coherent pattern: all microarchitectures except Sandy Bridge fall through forward jumps.
We will use these architecture-dependent results to adapt the code snippets used in the following experiments, especially regarding prediction expectation and load buffer usage.

\begin{table}[t]
    \setlength{\extrarowheight}{5pt}
    \small
    \begin{tabularx}{\columnwidth}{|l|X|l|}\hline
        \normalsize\textbf{$\mu$-arch.}     & \normalsize \textbf{Model Name}                                   & \normalsize \textbf{Year} \\\hline
        Skylake         & Intel\textsuperscript{\textregistered} Core\textsuperscript{\texttrademark} i5-6200U @2.30GHz & 2015 \\\hline
        Haswell         & Intel\textsuperscript{\textregistered} Core\textsuperscript{\texttrademark} i5-4690 @3.50GHz & 2013 \\\hline
        Sandy Bridge    & Intel\textsuperscript{\textregistered} Xeon\textsuperscript{\textregistered} E5-2430 v2 @2.50GHz     & 2011 \\\hline        
        Nehalem         & Intel\textsuperscript{\textregistered} Xeon\textsuperscript{\textregistered} L5520  @2.27GHz         & 2008 \\\hline        
        Prescott        & Intel\textsuperscript{\textregistered} Xeon\textsuperscript{\texttrademark} 3.20GHz                & 2004 \\\hline 
    \end{tabularx}
    \caption{List of Intel's microarchitectures tested.}
    \label{tbl:uarchs-tested}
\end{table}


For Kernel ASLR derandomization we use both our attack techniques from \Cref{sec:double-speculation-kernel} and \Cref{sec:buffer-exhaustion}.
In the following section we will present our findings.

\subsection{Measuring the Execution Time}
\label{sec:measure-mem-acccess-time}
In \Cref{sec:attack} we have described ways to identify whether a page is mapped or not.
All feedback channels we presented relied on timing differences due to memory caching.
To measure such time differences between cached and non-cached memory accesses, we use hardware timestamp counters (\asm{rdtsc} and \asm{rdtscp}).
Such counters deliver higher than nanosecond precision and are sufficient to differentiate the two cases.
However, as these instructions could be reordered during the execution on a CPU, care needs to be taken to get the measurements as precise as possible.
To this end, we resort to serializing instructions that avoid instruction reordering (in particular, we use \asm{cpuid}).
The code for measuring a memory access at address \texttt{U} is shown below:
\begin{asmcode}[label=lst:measure-mem-acc-rdtscp,caption=Measuring access time via timestamp counters.]
  cpuid               ; serializing point
  rdtsc               ; start measuring
  mov     r11, rax    ; store measurement
  mov     rax, [U]    ; access memory
  rdtscp              ; end measuring
  mov     r12, rax    ; store measurement
  cpuid               ; serializing point
  sub     r12, r11    ; r12 - time tiff
\end{asmcode}

\noindent
The code in \Cref{lst:measure-mem-acc-rdtscp} performs the following steps (by line): 
\begin{enumerate}
    \item \asm{cpuid} waits for all preceding instructions to finish before it gets executed. 
    This introduces a serializing point to ensure that no instruction before it will get executed after.
    
    \item \asm{rdtsc} reads the initial counter value, which will be returned in \asm{rax:rdx} register pair.
    
    \item We are interested in smaller values that can fit in 32 bits and thus only store the lower part of it, i.e., \asm{eax}, in \asm{r11}.
    
    \item We read the data at address \asm{U}. This is the access for which we want to measure the time.
    
    \item For the second measurement, we use the \asm{rdtscp} instruction to get the final timestamp counter.
    \asm{rdtscp} is specially designed to read the timestamp counter only after all memory operations before it are done, which exactly matches our needs.
    
    \item We store the new counter value into \asm{r12}.
    
    \item To be sure that no consecutive instructions will be scheduled together with \asm{rdtscp} instruction, we use another serializing point (\asm{cpuid}).
    
    \item Using the two timestamp values in \asm{r12} and \asm{r11}, we subtract them to get the difference.
\end{enumerate}

This code allows us to measure only the time needed for memory access, without adding too much noise.
For older CPUs that do not support \asm{rdtscp} yet, we resort to using \asm{rdtsc} for the second measurement as well and additionally preceding it with \asm{cpuid}, so that \asm{rdtsc} is not scheduled together with the memory access that we want to measure.
By doing so we introduce another instruction (\asm{cpuid}) in between two measurement points.
This will add some overhead to the returned values.
Nevertheless, the overhead will be included in both cached and non-cached memory accesses, and thus the difference will remain the same.

\begin{table}[t]
\begin{minipage}{\columnwidth}  
    \setlength{\extrarowheight}{5pt}
    \small
    \begin{tabularx}{\columnwidth}{|l|l|X|X|}\hline
        \normalsize\textbf{$\mu$-arch.} & \normalsize \textbf{FJ\textsuperscript{1}} & \normalsize \textbf{LB\textsuperscript{2} Entries} & \textbf{PL\textsuperscript{3}} \\\hline
        Skylake         & N & 72 & 40 \\\hline
        Haswell         & N & 72 & 32 \\\hline
        S-Bridge        & N & 64 & 32 \\\hline        
        Nehalem         & T & 48 & 11 \\\hline        
        Prescott        & N & -- & 19 \\\hline 
    \end{tabularx}
    \caption{Prediction behavior and characteristics.}
    \footnotesize\textsuperscript{1}Forward Jump: \textbf{T}aken, \textbf{N}ot taken, \textsuperscript{2}Load Buffer, \textsuperscript{3}Parallel Loads
    \hrule
    \label{tbl:uarchs-properties}
\end{minipage}
\noindent\hrulefill 
\end{table}

\subsection{Differentiating Between Mapped/Unmapped Kernel Pages}

\begin{figure*}[ht]
	\begin{subfigure}{.2\textwidth}
		\centering
		\includegraphics[width=1.0\linewidth]{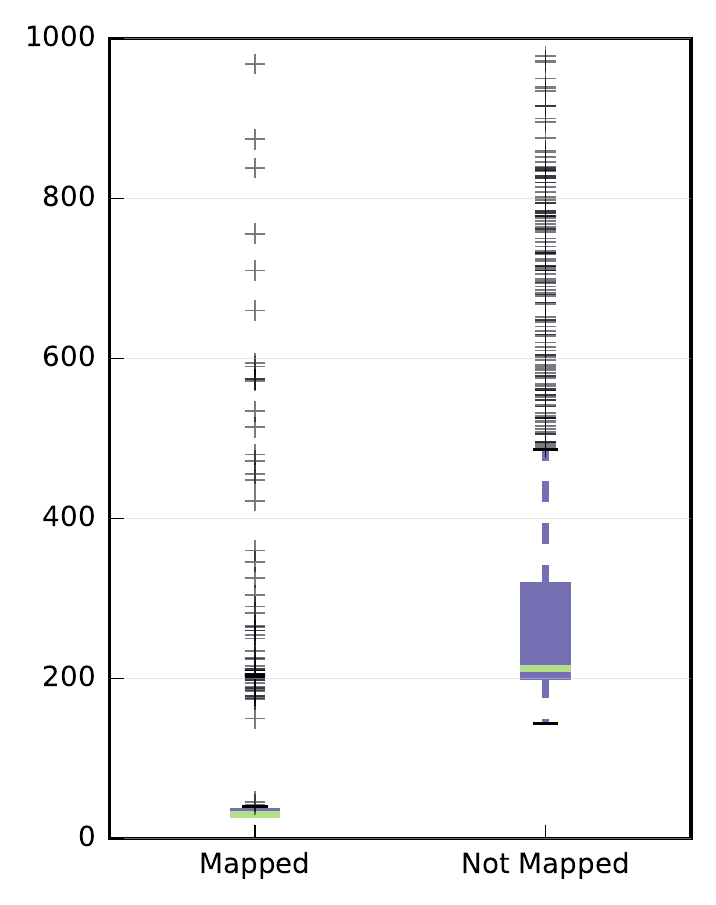}
		\caption{Skylake}
		\label{fig:boxplot_skylake}
	\end{subfigure}%
	\begin{subfigure}{.2\textwidth}
		\centering
		\includegraphics[width=1.0\linewidth]{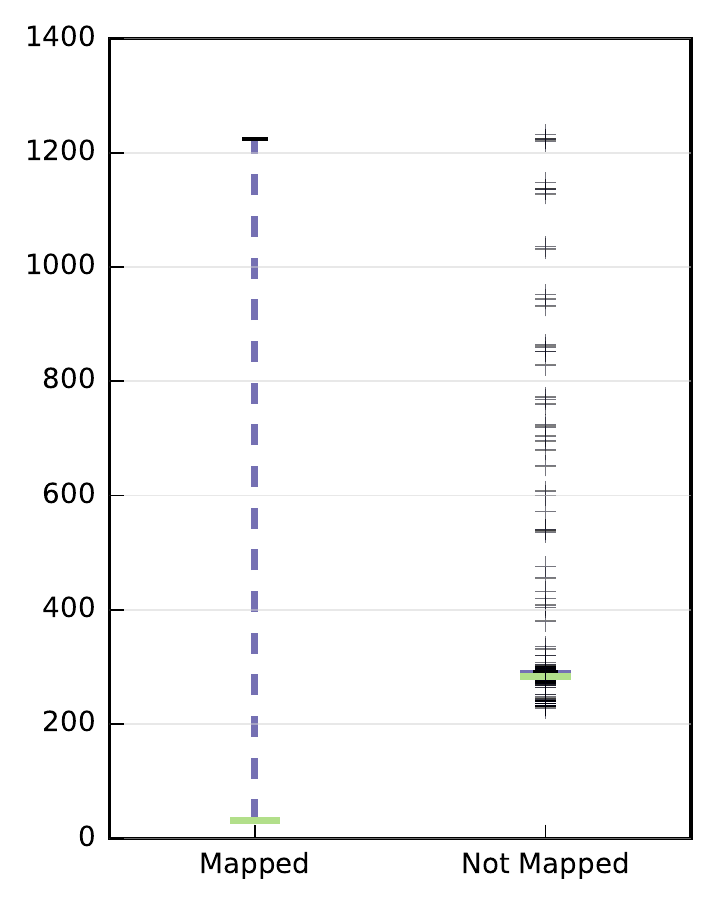}
		\caption{Haswell}
		\label{fig:boxplot_haswell}
	\end{subfigure}%
	\begin{subfigure}{.2\textwidth}
		\centering
		\includegraphics[width=1.0\linewidth]{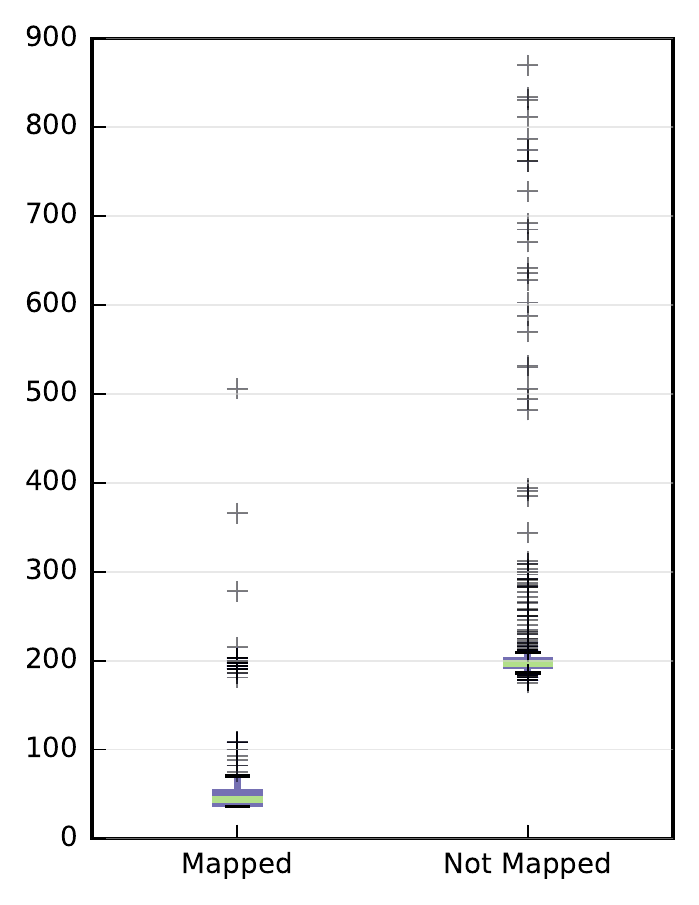}
		\caption{Sandy Bridge}
		\label{fig:boxplot_sandy}
	\end{subfigure}%
	\begin{subfigure}{.2\textwidth}
		\centering
		\includegraphics[width=1.0\linewidth]{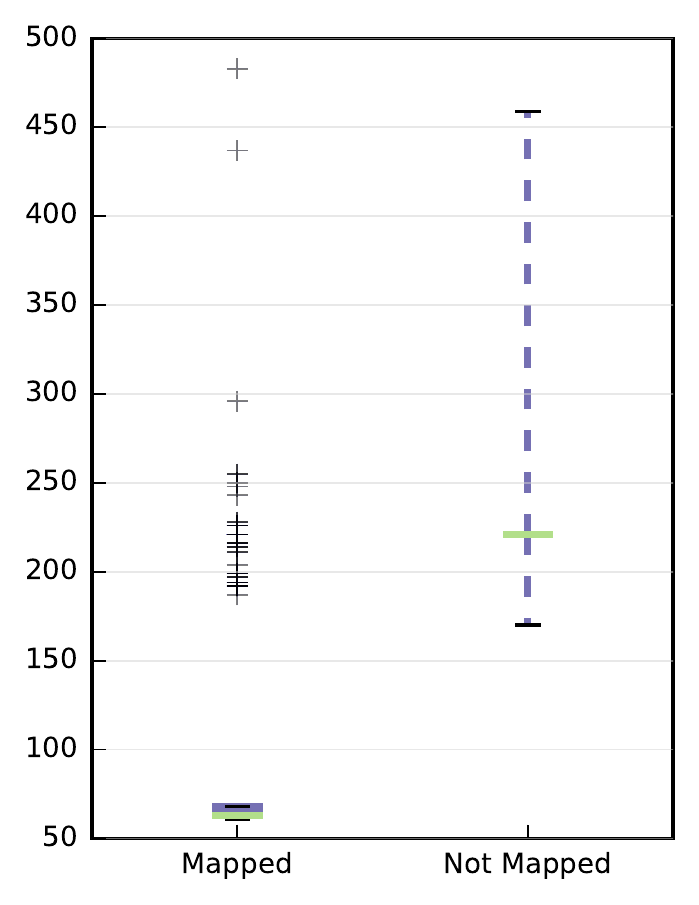}
		\caption{Nehalem}
		\label{fig:boxplot_nehalem}
	\end{subfigure}%
	\begin{subfigure}{.2\textwidth}
		\centering
		\includegraphics[width=1.0\linewidth]{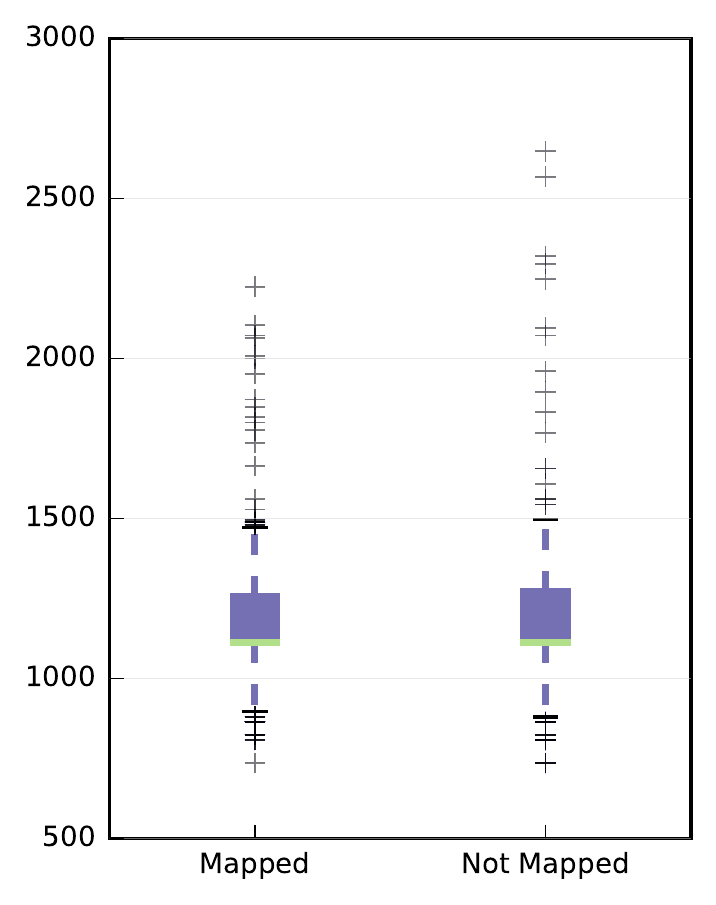}
		\caption{Prescott}
		\label{fig:boxplot_prescott}
	\end{subfigure}%
	\caption{Results of the caching side channel. If left (mapped) and right (not mapped) deviate, feedback is reliable.}
	\label{fig:march-cache-timings}
\end{figure*}

At this point we have all the prerequisites to measure timing differences accurately.
We will now use this methodology to evaluate the feedback channels as proposed in \Cref{sec:backreport} and \Cref{sec:probing-feedback}.
We will first focus on the two feedback channels that cache a specific user-mode memory address.
Given lack of conceptual differences, we will treat them equally in this section and benchmark the timing differences between accessing a cached (mapped kernel page) and non-cached (non-mapped kernel page) address using the two-level speculation method proposed in \Cref{sec:caching_feedback}.
\Cref{fig:march-cache-timings} summarizes the timings, more specifically their minimum, median, and average values over 1000 runs.
Given that we use \asm{rdtsc} for measurements, the timings approximate the number of cycles\footnote{http://www.forwardscattering.org/post/15} for corresponding memory accesses.
As we want to see whether memory is cached or not, we will later resort to the more reliable minimum values.

\Cref{fig:march-cache-timings} clearly shows timing differences between mapped and non-mapped pages for most architectures.
Two architectures are worth mentioning in particular, as their behavior differed from the others.
In Nehalem, accessing a non-mapped memory page does not stall the dependent instructions in the pipeline; instead it would also return 0 as a value.
Therefore, the technique defined in \Cref{sec:double-speculation-kernel} (two-level speculation in \Cref{lst:double-speculation-kernel} or dependent load in \Cref{lst:dependency-load-kernel}) will not work there.
However, resource exhaustion of architectural buffers (\Cref{sec:buffer-exhaustion}) works seamlessly and we report these numbers instead.
In particular, following \Cref{tbl:uarchs-properties}, we used between 11 and 47 load instructions to the probed address followed by a caching instruction, as Nehalem features 11 parallel cache miss and 47 load buffer entries, respectively.

Prescott also prevents the first approach, as both faulting addresses stall the pipeline and therefore the caching instruction is not executed at all.
Similarly, the resource exhaustion technique (\Cref{sec:buffer-exhaustion}) also fails.
Regardless of the type of fault (i.e., page fault or access violation), consecutive load instructions are stalled after 19 faulty ones.

The flushing-based feedback channel turned out to be less reliable than caching-based feedback, as the minimum measurements did not show any measurable difference regardless of the branch condition.
In the following, we will thus rely on the caching-based feedback channels that have shown to be accurate.
Having said this, we observed significant timing differences between confirmed-but-halted speculative execution (via \asm{hlt}) and mispredicted execution.
For example, on Sandy Bridge, the median of 1000 measurements differed in 16 cycles, and also the average execution time on Nehalem was significantly slower for non-mapped pages.
We refer the interested reader to \Cref{appendix:2ndfeedback} to see how and on which architectures one could use flushing-based feedback.

\subsection{Breaking Linux KASLR}
\label{sec:find-kernel-image}
By now, we can successfully identify whether a single kernel address space memory page is mapped, using any of our proposed approaches with equal success.
The dependent load technique (as shown in \Cref{lst:dependency-load-kernel}) is by far the simplest means of doing so, considering that it uses only two instructions.
We will thus use this test to derandomize KASLR on 64-bit Linux (Ubuntu 16.04 LTS; kernel 4.13.0) running on a 4-core Haswell CPU with 32GB RAM.
Kernel ASLR is enabled in the OS by default, which, as described in \Cref{sec:background-KASLR}, randomizes the location of the kernel image in the range from \texttt{0xffffffff\textbf{800}00000} to \texttt{0xffffffff\textbf{c00}00000}.
Given that the starting address of the image is aligned to 2MiB pages this results in 9 bits of randomness.

For the derandomization attack, we created an executable, which takes the probe address as an input and outputs the time it takes to access the user-mode memory address (i.e., the one that is being cached in the case of the mapped kernel page).
To reduce the possible noise, we run the program twice per probed address and look at the minimum duration of the two runs.
We use a single-threaded process to run the experiment.
Increasing the number of parallel processes is possible, but would introduce noise because of hyper-threading, which would require more than two tests per probed address.
According to our experiments, a single process running each probe twice showed the best compromise between efficiency and accuracy.
Using the program as an oracle, we probe each possible address in the KASLR memory range with 2MiB increments (which corresponds to Linux KASLR's alignment size).
In principle, this could be further improved with larger increments that take into account the size of the kernel image, or by using a binary (instead of linear) search.
That is, note that we did not strive towards decreasing the search time, in order to find a generic solution that works with any kernel image size (even small ones).
Still, using this na\"ive search, we reliably find every mapped 2MiB kernel page in 2.63 seconds on average.

We use the same approach to search for kernel modules, but modified the search area, given that the possible start address of modules is between \texttt{0xffffffffc0\textbf{000}000} and \texttt{0xffffffffc0\textbf{c00}000} (page-aligned).
This gives us 3072 possible places where modules can be allocated ($\sim$11 bits of entropy).
Running the above experiment for modules with the modified parameters found all allocated pages for modules reliably in 14.70 seconds on average.

To evaluate the accuracy of this scheme, we ran the program 1000 times. 
We use the list of kernel pages in \texttt{/sys/kernel/debug/kernel\_page\_tables} as a ground truth to detect possible false positives and false negatives.
Out of 1000 runs, our program always correctly identified 11 out of 11 kernel image pages.
As for detecting loaded modules, our program successfully identified all 1316 mapped module pages in 968 out of 1000 runs.
For the remaining 32 runs it only missed a single mapped page.
Throughout the test run, our program did not report any false positives.

\subsection{Breaking Linux KASLR in a VM}
\label{sec:virtual-derandomize-kaslr}
So far we have constrained the proposed attacks to physical hardware, yet an attacker might be interested to abuse similar techniques in a virtualized environments.
A core driver for the recent increase of virtualization was the CPUs' virtualization support, which made the execution more efficient.
One of the features that hardware support nowadays is extended page tables (EPT), which adds another layer of translation to page tables in order to translate guest physical addresses into host physical pages.
Requiring no hypervisor intervention to translate virtual addresses, we can use EPTs to mount our attack even on a VM.
In our next experiment, we carry out our measurements on virtualized 64-bit Linux (Ubuntu 16.04 LTS; kernel 4.14.0) running on a 4-core Haswell CPU with 8GB of virtual RAM.
For virtualization we use VirtualBox (Version 5.1.30) with hardware acceleration enabled (the default).

We ran the same measurements as we did for physical machines (\Cref{sec:find-kernel-image}).
As expected, a virtualized environment reduced the efficiency of our approach.
Keeping the same parameters (i.e., 2 trials per probed address on a single core), the measurements took on average 2.51 seconds and 17.78 seconds to find mapped pages for the kernel image and loaded modules, respectively.
Further, the precision of the approach was decreased, missing 47 out of 528 pages on average (min/max being 26 and 72 respectively, and median being 47) for modules, and missing 2 out of 11 pages on average for the kernel image.

To cope with this reduced accuracy, in a second experiment, we doubled the number of trials per probed address (i.e., 4 trials instead of 2).
This significantly improved the precision of our approach; however, it also doubled the execution time.
With increased trials, it takes on average 5.66 seconds to find the kernel image and 33.80 seconds to find the loaded modules.
As for the precision of the attack, we observed on average 5 missed pages for modules (min/max being 0 and 52, respectively, and median being 4) and 1 missed page of the kernel image.
Note, however, that despite the missed pages, it is trivial for an attacker to correctly identify the area of the kernel image if the two missing pages do not constitute pages at the edge of the image---which, given the overall size of the region found, can be easily verified.
Additionally, one can probe for neighboring pages after the initial scan to find possible false negatives, or simply repeat measurements.

\subsection{Breaking Windows KASLR}
\label{sec:find-kernel-image-Windows}
After we have demonstrated how security guarantees of Linux KASLR are undermined with speculative execution, we will now turn to the Windows OS.
We chose a Skylake CPU and Windows~10 (version 1709) for our evaluation.
Despite being OS agnostic, we still had to adjust our test program for Windows.
More specifically, we have to work around user-land ASLR implementation differences between Windows and Linux.
In contrast to Linux, which randomizes executables at every run, Windows randomizes them using a fixed seed determined at boot time.
Consequently, running the same program multiple times does not randomize base addresses anymore, which we used to evade dynamic prediction.
To solve the problem, we created a single executable probing multiple addresses one after another in consecutive speculative memory accesses, derandomizing the address space in a single run.
As a side effect, this drastically improved the performance of our test program.
Similar to our attack against Linux Kernel~\Cref{sec:find-kernel-image}, we will also run this program as a single process, this time using just one probe per address.

The second challenge in Windows comes from KASLR itself.
We observed that the kernel image is always loaded together with the hardware abstraction layer (HAL) module.
However, the loading order is randomized (e.g., Kernel$\rightarrow$HAL or HAL$\rightarrow$Kernel).
Additionally, although both the kernel image and HAL are allocated on consecutive large pages (pages that are 2MiB large instead of 4KiB), their entry points are randomized inside the large page and can start at any 4KiB boundary.
Given that we can only probe for mapped pages, this attack only allows us find the base address of the large page containing the kernel image and HAL, and not their actual addresses.
However, this still removes 18 bits of KASLR randomness and can be used in combination with other techniques (e.g., those discussed in \Cref{sec:relatedwork}) to break the remaining 9~bits (i.e., the 4KiB-aligned kernel offset).

According to \Cref{sec:background-KASLR}, we scan the kernel address range from \texttt{0xfffff80000000000} to \texttt{0xfffff88000000000}.
Given the alignment of large pages, we step through the memory in 2MiB increments, giving us in total ‭262144‬ tries.
We verified that only the kernel image would allocate five consecutive 2MiB pages and use this as its fingerprint.
Running the experiment 100 times showed that the kernel image area in Windows can be found in under a second (0.55s) on average\footnote{Note that the additional overhead in Linux was due to loading and removing the executable for each memory probe, which could also be heavily optimized.}.
Out of 1000 runs, all 5 pages were found 74 times, while missing a single page 15 times and missing two pages 11 times.
Throughout the test run, we did not see any false positives.

\section{Discussion}
\label{sec:discussion}

\subsection{Security Implications of Speculative Execution}
\label{sec:usermode}
We believe that showing how one can derandomize KASLR is just the tip of the iceberg of security implications that speculative execution may have.
Executing arbitrary code on the CPU without consequences (such as segmentation faults), and being able to report results back is a powerful tool, which could have many different use cases.
One such case is using the arbitrary memory read technique (\Cref{sec:arbitrary-memory-read}) from inside a restricted environment that sandboxes memory accesses using bounds checking.
For example, consider a JavaScript execution environment in Web browsers that checks if memory accesses in attacker-controlled code target outside of a well-defined safe region (and thus should be rejected).
Given that the majority of modern browsers compile JavaScript into native code, executing arbitrary code on a CPU is feasible even from web applications.
Therefore, to gain arbitrary memory reads from inside JavaScript, one has to generate native code similar to that described in \Cref{sec:arbitrary-memory-read}.
Such an attack could be simplified by using WebAssembly, which allows compilation of C/C++ programs into a WebAssembly byte code, which itself will be compiled into native code by browsers.
We leave an evaluation of such attacks open for future work and hope that our analysis on speculative execution will foster new research to understand the full scope of the problem.

\subsection{Possible Mitigations}
We now turn to possible mitigation techniques that can be applied to make systems resistant to our attack.

\textbf{Hardware Modifications:}
First, we discuss approaches that require hardware modifications.
The naive approach would disable speculative execution altogether, which, however, would drastically reduce the performance of the CPU---every branch instruction would stall the execution pipeline.
A more reasonable approach would be to stall the execution when a privilege fault occurs.
This successfully removes the information leak that we use to distinguish between mapped and non-mapped pages.
However, this technique does not get rid of the general problem of side channels in speculative execution.
We have already provided two feedback channels in \Cref{sec:backreport}: (i) measuring the complete execution time and detecting difference of rollback complexity in case of mispredictions, (ii) and caching user-space memory.
To defeat the former feedback channel, one would need to unify execution times for micro-OPs.
Disabling the latter would require forbidding memory loads in speculative execution.
However, given that memory accesses are already a bottleneck for modern CPUs, blocking their speculation will incur a significant slowdown.
Note that merely disabling \emph{nested} speculation would not remedy the problem, as one could cache two different memory pages based on the value to leak in first-level speculation.
Summarizing, while mitigation in hardware is possible, added rigorous security would imply significant performance degradation.

\textbf{Attack-Specific Defenses:}
Although the problem originates from hardware, some attack-specific mitigations can also be applied in software.
One such mitigation against our KASLR attack is stronger kernel/user space isolation, which has been proposed multiple times~\cite{shadow_kernels,tale_of_two_kernels,ret2dir_rethink_kernel_isolation,netsted_kernels}.
The basic idea is to separate kernel and user address spaces into different virtual spaces.
This would remove kernel addresses entirely from user-level page tables, thus making them inaccessible.
The downside to this approach is the overhead caused by flushing translation lookaside buffers (TLBs) on every context switch from user to kernel mode.
Alternatively, one could try to increase the entropy of KASLR, which however would only slow down the attack, and efficient binary search techniques would reduce the search effort to logarithmic complexity.
Also, while any of this would mitigate leaks from kernel into user mode, as we have discussed in the previous subsection, even pure user-mode attacks can be of concern.

It seems more promising to hide the kernel content rather than just its existence.
One such solution was recently implemented in OpenBSD and is dubbed Kernel Address Randomized Link (KARL).
Instead of randomizing the base address of the kernel image, KARL randomizes the kernel's content (similar to proposed fine-grained randomization schemes in user space~\cite{ASLP,STG,STIR}).
At every system startup a new kernel image will be linked together by combining object files in a random order.
Having a completely different layout at each boot forbids the attacker to predict the addresses of required code or data pieces.
KARL thus withstands our proposed derandomization attack, as an attacker can only learn where the kernel is mapped, but not how it is laid out.

\textbf{Compiler-Assisted Solutions:}
An interesting defense opens up in a setting where the attacker does not control the compiler that generates the measurement code.
For example, consider a setting where an adversary can specify code that WebAssembly ultimately compiles into a byte code.
Assume the attacker aims to read beyond a critical memory check via speculative execution.
The compiler, assuming knowledge of such critical conditionals (e.g., by code annotations) could then create specific branches that are \emph{guarded} against misuse.
\Cref{lst:guarded-if} shows an example of such a guarded conditional in a BTFNT setting for the compiled code snippet \texttt{if (rax == somevar) \{...\} else \{...\}}.
To prevent speculative execution, a guard injects a backward jump that is \emph{never} taken (line~\#4) in normal execution, but will be predicted---effectively creating an infinite loop executed in speculative execution.
This way an attacker has no possibility to inject code in the fall-through case that is executed speculatively.

\begin{asmcode}[label=lst:guarded-if,caption={A guarded conditional hinders static misprediction, assuming BTFNT in this example.}]
  cmp rax, [somevar]
  je Equal
FallThrough1:
  je FallThrough1 ; GUARD: backward jump
                  ;  that is taken in BTFNT
  ... ; code in else {} branch
  jmp Exit
Equal:
    ... ; code in if {} branch
\end{asmcode}

\subsection{Future Work}
\textbf{Affected Vendors and ISAs}:
The focus of our experiments was Intel x86 CPUs.
We have not tried to apply our speculative execution-based attacks against other architectures, but also have not seen any fundamental reasons why this should not be possible.
For example, the CPUs of other CISC vendors like AMD also offer speculative execution engines that would open up similar side channels.
In principle, even RISC processors like ARM-based CPUs feature speculative execution, such as Cortex-R processors~\cite{cortex-R7-tech-ref} or the Cortex-A57 in LG Nexus 5X smartphone~\cite{cortex-A57-nexus5x-spec-exec}.
Having said this, the inner workings of CPUs might differ significantly.
For example, if a CPU does not feature nested speculation, the presented side channels require adaptations. 
With this work, we have proposed several automated tests that identify whether a certain CPU is susceptible to speculation-based side channels.
Using this test suite, we aim to broaden our analyses to other architectures in the future.

\textbf{User-Mode Attacks}:
We have already conceptually described that user mode processes might also be at risk due to speculative execution (\Cref{sec:usermode}).
In immediate future work, we plan to test whether sandboxed environments that allow attacker-controlled code to be JIT-compiled (in particular browsers) can be abused to read out-of-bounds.
Such a proof-of-concept will be a significant research and engineering effort on its own and is thus out of scope for this paper.

\vspace{1em}
\section{Related Work}
\label{sec:relatedwork}

In the following, we list some of the works that are related to our approach.
This includes recent works targeting KASLR, but also side channels in general.
We highlight that none of these works has proposed to use speculative execution for their side channels.
In contrast, our work is the first to show how (i) speculation can be reliably abused to execute attacker-controlled code  that is never executed without speculation, (ii) that such speculative execution is \emph{not} free of side effects, as assumed so far, and (iii) that even modern KASLR implementations using a high entropy (such as in Windows) are undermined when facing speculation.

\subsection{Using Hardware to Challenge (K)ASLR}
\etal{Hund}~\cite{kaslr_break_practicaltiming} propose timing-based side channel attacks against KASLR.
They describe three different scenarios, Cache Probing, Double Page Fault, and Cache Preloading,  in which the attacker is able to leverage side-channels in the hardware to derandomize KASLR.
The general observation of these attacks is that, even though the memory accesses in kernel space result in privilege faults, their corresponding cache/translation entries are still stored and allow faster consecutive accesses.

The translation cache side channel was further improved by \etal{Jang}~\cite{kaslr_break_tsx} by using Intel's TSX (Transactional Synchronization Extensions).
TSX allows handling faulty memory accesses without OS intervention, reducing measurement noise.
The authors managed map the whole kernel space, as well as distinguish their executable privileges.

Instead of accessing privileged memory pages directly, \etal{Gruss}~\cite{kaslr_break_prefetch} suggest using prefetch instructions.
Some of the advantages of prefetch instructions are that they do not cause page faults and ignore privilege checks, however, they still go through the same translation/lookup as regular memory access would, and thus leave a measurable trace behind.

Another KASLR derandomization attack, from \etal{Evtyushkin}~\cite{kaslr_break_branchpredictor}, is more related to our approach, in that it also used branch predictors.
However, instead of abusing speculative execution to run arbitrary code without consequences, the authors try to cause collisions in branch target buffers (BTB) and thus leak the lower addresses of randomized pages.

Also user-space ASLR has been challenged by others.
\etal{Gras}~\cite{alsr_break_AnC} use MMU features to derandomize ASLR to mount an attack by executing JavaScript code on a victim's browsers.
\etal{Bosman}~\cite{aslr_break_deduplications} use the memory deduplication feature in operating systems to leak memory pointers from JavaScript, thus also resulting in successful ASLR derandomization (although the consequences of the complete attack were more significant than breaking ASLR).
Similarly, \etal{Barresi}~\cite{aslr_break_dedup_VM} use memory deduplication in virtual machine monitors to leak the address space layout of neighbor VMs.

\subsection{CPU-Related Side Channels}

The need to run multiple processes on the same machine creates a requirement to share limited resources among all of them.
This opens an opportunity for an adversarial process to leak sensitive information about the environment by looking at the utilization of those resources.
For example, instruction caches have been used to leak the information about the execution trace of co-existing programs on the same execution core~\cite{yet_another_microarch_attack_icache,new_results_icache,rsa_vuln_icache,improve_rsa_attack_icache}, or even across different VMs on the same machine~\cite{cross_vm_icache}, allowing the attacker to reconstruct sensitive information, e.g., private keys of cryptographic protocols.

In contrast to instruction caches, where the attacker recovers execution trace of the program, data cache attacks can reveal the patterns of data accesses of neighboring processes and thus reveal the underlying operation.
These types of attacks have been shown to be successful in reconstructing different cryptographic secrets, such as AES keys~\cite{efficient_cache_attacks_dcache}.
Similarly, a last-level data cache side channel can be used reconstruct private keys~\cite{cache_games_dcache}, or leak sensitive information across VMs, e.g., keystroke timings to snoop user-typed passwords in SSH~\cite{hey_get_off_my_cloud}.

One of the techniques of leaking cache usage information in x86 is Flush\&Reload attack~\cite{flush+reload}.
In this attack, the attacker issues the \asm{clflush} instruction to flush victim's cache lines.
Measuring the same cache lines later will reveal if it has been accessed.
However, Flush\&Reload assumes that the attacker has access to victim's memory pages (e.g., via shared pages).
Prime\&Probe~\cite{last_level_cache_practical}, in contrast, does not require shared memory regions, and instead relies on cache collisions on shared caches between the attacker and the victim.

\vspace{1em}
\section{Conclusion}
\label{sec:conclusion}
Speculative execution has long been ignored by the security community, although it bares critical threats that undermine important assumptions of existing defensive schemes.
We are the first to provide a comprehensive overview of the security implications.
Such an understanding is crucial for security researchers to understand the guarantees of existing solutions.
Our adversarial setting may seem contrived compared to usual threat models, as we assume that the attacker has influence on the code.
However, we argue that the proposed side channels will highly influence existing defenses, such as conditionals that constrain unsafe code to certain memory regions---especially in the realm of JIT compilation like in browsers.
The recent rise of KASLR techniques demonstrates that abusing speculative execution is clearly a novel type of attack class that needs further considerations.
\vspace{3em}

\Urlmuskip=0mu plus 1mu\relax
\begin{small}
\bibliographystyle{IEEEtranS}
\bibliography{paper}
\end{small}

\pagebreak
\section{Appendix}

\begin{figure}[h]
	\centering
	\begin{subfigure}{.33\columnwidth}
		\centering
		\includegraphics[width=1.0\linewidth]{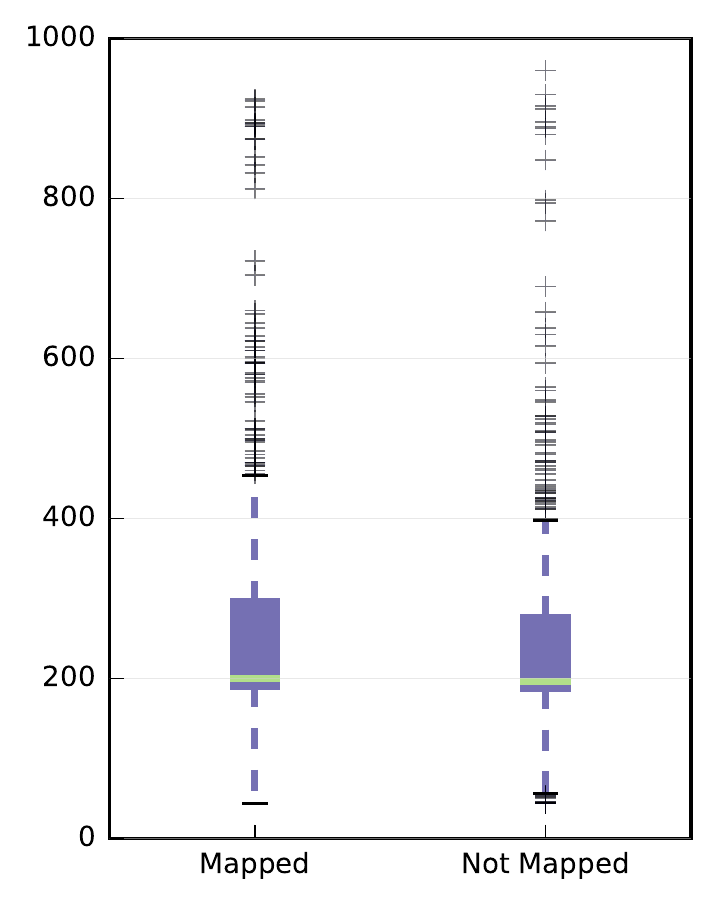}
		\caption{Skylake}
		\label{fig:all_overheads_skylake}
	\end{subfigure}%
	\begin{subfigure}{.33\columnwidth}
		\centering
		\includegraphics[width=1.0\linewidth]{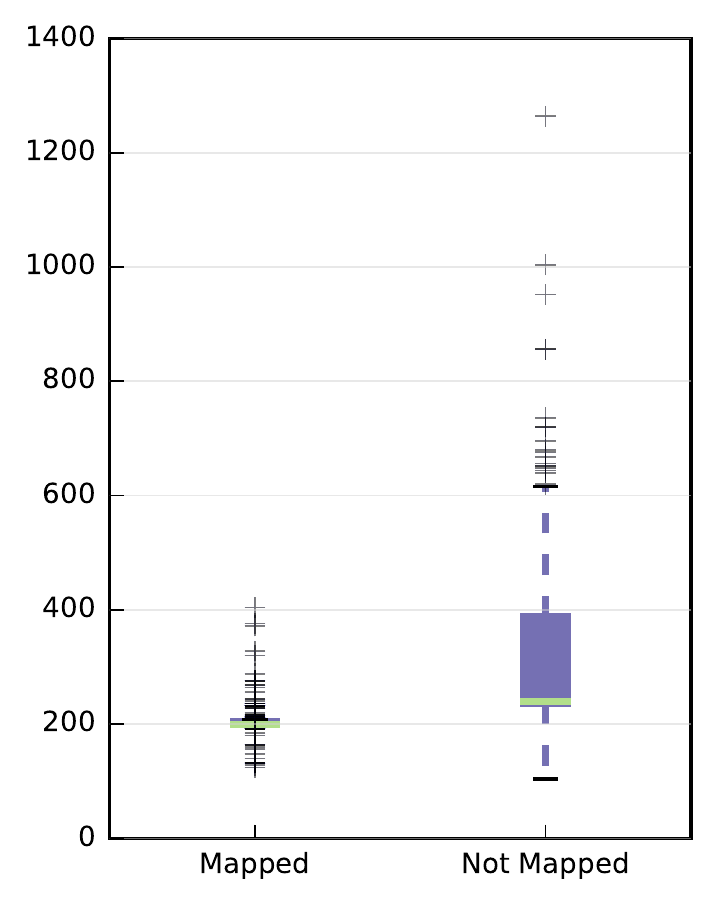}
		\caption{Haswell}
		\label{fig:all_overheads_haswell}
	\end{subfigure}%
	\begin{subfigure}{.33\columnwidth}
		\centering
		\includegraphics[width=1.0\linewidth]{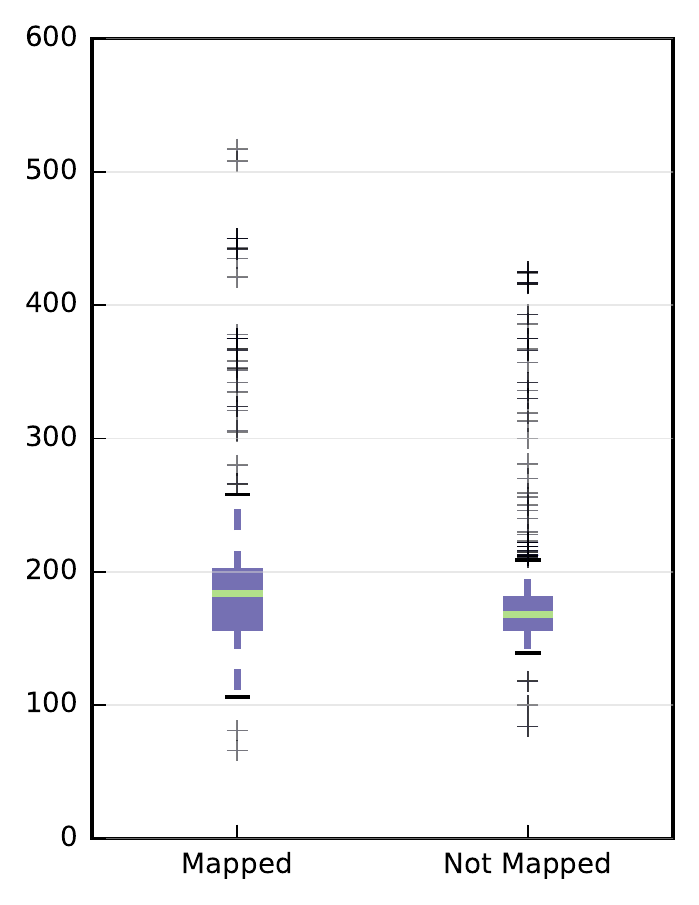}
		\caption{Sandy Bridge}
		\label{fig:all_overheads_sandy}
	\end{subfigure}\\%
	\begin{subfigure}{.33\columnwidth}
		\centering
		\includegraphics[width=1.0\linewidth]{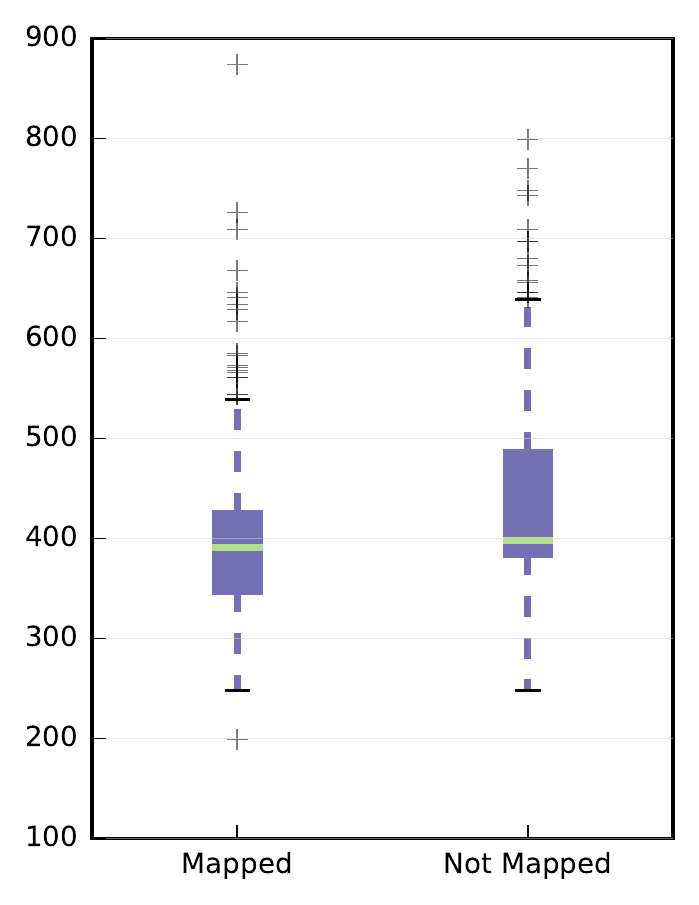}
		\caption{Nehalem}
		\label{fig:all_overheads_nehalem}
	\end{subfigure}%
	\begin{subfigure}{.33\columnwidth}
		\centering
		\includegraphics[width=1.0\linewidth]{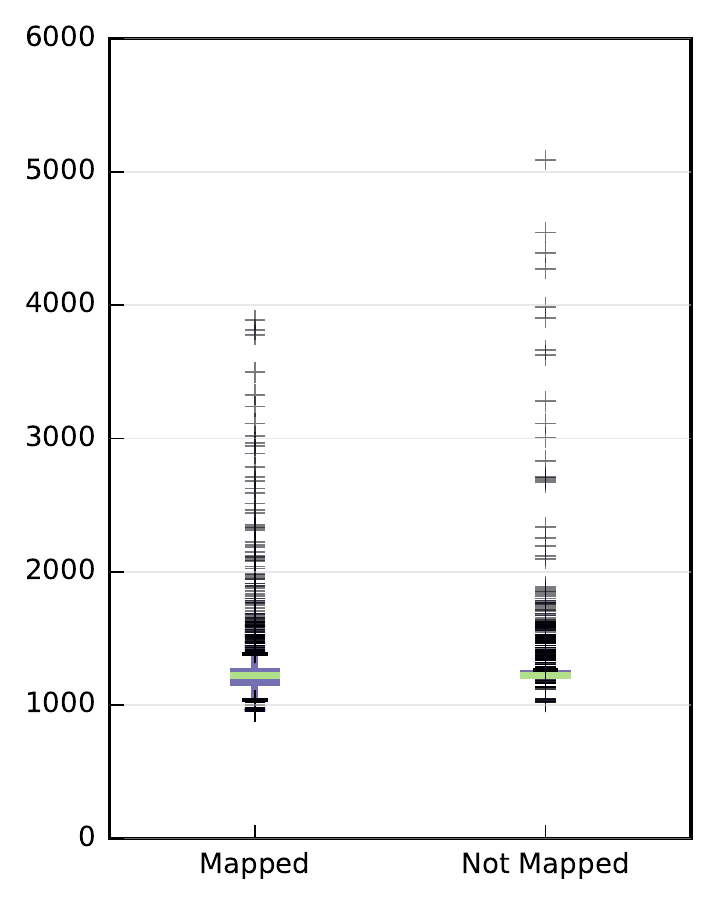}
		\caption{Prescott}
		\label{fig:all_overheads_prescott}
	\end{subfigure}%
	\caption{Results of the flushing side channel. If left (mapped) and right (not mapped) deviate, feedback is reliable.}
	\label{fig:2ndfeedback}
\end{figure}

\subsection{Flushing-Based Side Channel Analysis}
\label{appendix:2ndfeedback}
This section details evaluation results of our flushing-based side channel.
In \Cref{sec:evaluation}, we measured and compared the minimum number of cycles for cache accesses.
We cannot directly apply the same methodology to the execution time side channel.
First, we cannot wrap our timing measurements in two serializing instructions such as \asm{cpuid}, as this would allow the sequence of \asm{imul} instructions to complete---avoiding speculation.
Therefore, we removed the first call to \asm{cpuid}, and also included the entire speculative execution block in our measurements (instead of access to a memory address).

Second, we found that the flushing execution time differs significantly between repeated executions, frequently not showing any difference in the condition (e.g., page mapped or not).
This implies that the minimum execution time among several (again, we used 1000) executions cannot be used to leak a condition, but instead we have to inspect the value distributions, averages or median values.

To allow for a fair comparison between the feedback channels, we provide the flushing-based feedback measurement results for architectures in \Cref{fig:2ndfeedback}.
For two architectures, Prescott and Skylake, there is no significant measurable timing difference between the page being mapped or not.
We believe that Prescott fails for the same reason as for the other feedback channels.
But also Skylake, the most recent architecture in our measurements, does not yield a measurable difference in our concrete instantiation of instruction flushing.
While improved side channels of similar methodology (e.g., other instructions than \asm{hlt}) might still exist, this indicates that recent Intel CPUs have more stable execution times regardless of speculation.

The other three architectures, however, show significant differences that can reliably be confirmed when repeating the experiment.
To our surprise the results were not as coherent as expected.
Haswell (\Cref{fig:all_overheads_haswell}) and Nehalem (\Cref{fig:all_overheads_nehalem}), on average, showed faster executions for mapped pages.
In contrast, Sandy Bridge (\Cref{fig:all_overheads_sandy}) slows down executions (median) accessing mapped pages.
We cannot explain this deviation in detail, but speculate that the inner workings of the CPU's pipeline engine and its different branch prediction rollback algorithms lead to such characteristic behavior.
\end{document}